\documentclass[letterpaper, 10 pt, conference]{ieeeconf}

\IEEEoverridecommandlockouts                              

\usepackage{times}

\usepackage{multicol}
\usepackage[bookmarks=true]{hyperref}

\usepackage{mathtools}
\usepackage{amsthm}
\usepackage{amssymb, amsfonts}
\usepackage{cases}

\usepackage{enumerate}

\newcommand{\qD}{\dot{q}}
\newcommand{\qDD}{\ddot{q}}

\newcommand{\Dimq}{n}

\newcommand{\T}{\top}

\newcommand{\nc}{n}
\newcommand{\coll}{c}

\newcommand{\colvec}[2]{\bigl( \begin{smallmatrix}#1 \\ #2 \end{smallmatrix} \bigr)}

\newcommand{\id}[1][3]{{I}}
\newcommand{\zero}[2]{{0}}
\newcommand{\zeros}[2]{{0}}

\newcommand{\multeqi}[2]{\begin{IEEEeqnarraybox}[][#2]{#1}}
\newcommand{\multeqf}{\end{IEEEeqnarraybox}}

\newcommand{\systemi}[1][rCL]{\left\lbrace\begin{IEEEeqnarraybox}[][c]{#1}}
\newcommand{\systemf}{\end{IEEEeqnarraybox}\right.}

\newcommand{\eqni}[1][rCL]{\begin{IEEEeqnarray}{#1}}
\newcommand{\eqnf}{\end{IEEEeqnarray}}

\newcommand{\nneqni}[1][rCL]{\begin{IEEEeqnarray*}{#1}}
\newcommand{\nneqnf}{\end{IEEEeqnarray*}}

\newcommand{\pmatrixi}{\begin{pmatrix}}
\newcommand{\pmatrixf}{\end{pmatrix}}

\newcommand{\bmatrixi}{\begin{bmatrix}}
\newcommand{\bmatrixf}{\end{bmatrix}}

\newcommand{\smatrixi}{\left[\begin{smallmatrix}}
\newcommand{\smatrixf}{\end{smallmatrix}\right]}

\newcommand{\enumi}{\begin{enumerate}}
\newcommand{\enumf}{\end{enumerate}}

\newcommand{\enumri}{\begin{enumerate}\renewcommand{\theenumi}{\textit{\roman{enumi}}}}
\newcommand{\enumrf}{\end{enumerate}}

\newcommand{\mytheorem}[2]{%
\newtheorem{t#2}{#1}%
\newenvironment{#2}{\begin{t#2}}{\end{t#2}}}

\theoremstyle{plain}

\mytheorem{Theorem}{theorem}
\mytheorem{Proposition}{proposition}
\mytheorem{Lemma}{lemma}
\mytheorem{Corollary}{corollary}
\mytheorem{Assumption}{assumption}
\mytheorem{Definition}{definition}
\mytheorem{Property}{property}

\usepackage{graphicx}
\usepackage{subcaption}
\usepackage{paralist}
\usepackage{gensymb}
\DeclareGraphicsExtensions{.pdf,.png,.jpg}

\DeclareMathOperator{\tr}{tr}

\hypersetup{pdfinfo={
   Author={Francesco Romano, Daniele Pucci, Francesco Nori},
   Title={Collocated Adaptive Control of Underactuated Mechanical Systems},
   CreationDate={D:20140125120000},
   Subject={Robotics},
   Keywords={Control theory;Robotics,Underactuated},
}}

\title{\LARGE \bf
Collocated Adaptive Control of \\ Underactuated Mechanical Systems}

\author{Francesco Romano and Daniele Pucci and Francesco Nori$^{1}$
\thanks{*This paper was supported by the FP7 EU projects CoDyCo (No. 600716 ICT 2011.2.1 Cognitive Systems and Robotics), and Koroibot (No. 611909 ICT-2013.2.1 Cognitive Systems and Robotics).}%
\thanks{$^{1}$All authors belong to the Department of Robotics, Brain and Cognitive Sciences,
        Italian Institute of Technology, Via Morego 30, Genoa, Italy
        {\tt\small name.surname@iit.it}. }%
}

\begin{document}

\maketitle
\thispagestyle{empty}
\pagestyle{empty}

\begin{abstract}
Collocated adaptive control of underactuated systems is still a main concern for the control community,
all the more so because the collocated dynamics is no longer linear with respect to the constant base parameters. 
This work extends and encompasses the well known adaptive control result for fully actuated mechanical systems to the underactuated case. 
The key point is the introduction of a fictitious control input that allows us to consider the complete system dynamics, which is assumed to be linear 
with respect to the base parameters. 
Local stability and convergence of time varying reference trajectories for the collocated dynamics are demonstrated by using Lyapunov and Barbalat arguments. 
Simulation and experimental results on a two-link manipulator verify the soundness of the proposed approach.
\end{abstract}

\section{Introduction}

Nonlinear feedback control of underactuated mechanical systems is not new to the scientific community~\cite{Olfati-Saber2000, Hera2011, Liu2013}. Aircraft, underwater vehicles, and humanoid robots are 
only a few examples where the number of control inputs is fewer than the system's degrees of freedom, which characterizes the nature of an 
underactuated system \cite{Spong1998}.
Clearly, the lack of actuation along with model uncertainties significantly complexify the control problem associated with these systems. Given an open-chain mechanical
system, this work proposes
control strategies for a subset of the system's degrees of freedom by using estimates of its dynamical model. In the language of automatic control,
the laws presented in this paper fall into the category of \emph{adaptive control schemes}~\cite{Astrom1994}.

Underactuated mechanical systems arise specific issues when attempting the control of the complete set of degrees of freedom. 
Assuming that the system's desired configuration is feasible, the nature of a controller that asymptotically stabilizes this configuration is intimately related to the nature of the 
system itself. For instance, systems without potential terms in general forbid the existence of time-invariant feedback continuous stabilizers~\cite{Reyhanoglu1999}.
This claim, which follows from an application of Brockett's Theorem~\cite{Brockett83}, motivated the development of discontinuous and/or time-varying feedback stabilizers
for specific classes of systems~\cite{Luca1996, Park2009a, Ghommam2010}. Clearly, the complexity of the control problem reduces when attempting to stabilize
only a subset of the system's degrees of freedom.

In the specialized literature, several methods have been proposed to control a subset of the system's degrees of freedom. 
Inverse dynamics~\cite{Luca2000, Spong1998}, sliding mode~\cite{Santiesteban2008}, and energy based techniques~\cite{Spong1996} are among the main tools exploited by these works. 
The common denominator of these approaches is to partition the set of degrees of freedom into two subsets, usually referred to as
\emph{collocated} and \emph{noncollocated}. The former, whose cardinality equals the number of control inputs, 
contains the \emph{actuated} degrees of freedom. The latter accounts for the remaining
\emph{nonactuated} degrees of freedom -- see~\cite{Spong1998} for additional details. 
Then, the control objective is usually defined as the asymptotic stabilization of either set to desired values. 

To cope with model uncertainties, which may impair the effectiveness of the aforementioned approaches, adaptive control schemes have also been proposed. 
Adaptive stabilizations of the collocated and noncollocated set is achieved in~\cite{GuXu}. 
The main drawback of this approach is that the measurement of the system's acceleration is required by the feedback control action.
Leaving aside causality issues, this measurement may not be always available.

In the case of fully actuated mechanical systems, 
adaptive stabilization of time varying reference trajectories can be achieved~\cite{Slotine1988, Spong1990}.
The key assumption is that the system's dynamics can be expressed linearly with respect to a set of constant \emph{base parameters}.
The extension of these works to the underactuated case is not straightforward. As a matter of fact, the collocated dynamics is no longer linear with respect to the base parameters when
expressed independently of the noncollocated accelerations.

Assuming that the control objective is the asymptotic stabilization of the collocated degrees of freedom, the present paper basically extends~\cite{Slotine1988} to the underactuated case. 
The key point is the introduction of a fictitious control input that allows us to consider the complete system's dynamics, which is linear with respect to the base parameters.
No acceleration measurement is required by the proposed control laws. 

The paper is organized as follows. Section~\ref{sec:background} provides notation and background, and also proposes a slightly different formulation of \cite{Slotine1988} that simplifies both the presentation and the proof of the paper's contribution. The 
main control results are presented in Section~\ref{sec:main_result}. Validations of the approach are presented in
Section~\ref{sec:simulation}, first through simulations carried out with the simulator Gazebo, and then through 
experiments performed with
a two-link underactuated robot.  Remarks and perspectives conclude the paper.
\section{Background} 
\label{sec:background}

\subsection{Notation} 
\label{subsec:notation}
The following notation is used throughout the paper.
\begin{itemize}
 \item The set of real numbers is denoted by $\mathbb{R}$.
 \item Let $u$ and $v$ be two $n$-dimensional column vectors of real numbers, i.e. $u,v \in \mathbb{R}^n$, the inner product between them is 
 $x^\T y$, where ``$\T$'' stands for the transpose operator.
 \item Given a function of time $f(t)$, its time derivative is denoted by $\dot{f}(t)$. Given a function $f$ of several variables, its gradient
w.r.t. some of them, say $x$, is denoted as $ \partial_x f$.
 \item The euclidian norm of either a vector or a matrix of real numbers is denoted by $|\cdot |$.
\item $I_n \in \mathbb{R}^{n \times n}$ denotes the identity matrix of dimension $n$; $0_n \in \mathbb{R}^n$ denotes the zero vector of dimension $n$; $0_{n \times m} \in \mathbb{R}^{n \times m}$ denotes the zero matrix of dimension $n \times m$.
 \item The generalized coordinates characterizing the mechanical system are given by an $n$-dimensional vector of real numbers denoted as $q \in \mathbb{R} ^\Dimq$; its first 
 and second order time derivatives are indicated as $\qD$ and $\qDD$, respectively.
 \item $M(\cdot) \in \mathbb{R}^{n\times n}$, $C(\cdot) \in \mathbb{R}^{n\times n}$, and $g(\cdot) \in \mathbb{R}^{n}$ denote the inertia matrix, the Coriolis matrix, and the gravity torques obtained by applying Lagrange's formalism~\cite{Siciliano2008}.
\end{itemize}

 \subsection{System modeling and properties} 
 \label{subsec:modellingAndAssumptions}
 
 In light of the above notation, we assume that the application of Lagrange formulation yields a system's model of the following form:
 \begin{IEEEeqnarray}{RCL}
	 \label{eq:dynamics}
	 M(q,\pi)\qDD {+} C(q,\qD,\pi) \qD {+} g(q,\pi) {+} F_v(\pi) \qD {+} F(q,\qD,\pi)  = \tau \text{,} \IEEEeqnarraynumspace
\end{IEEEeqnarray}
 where $\pi \in \mathbb{R}^p$ is the vector of the (constant) system's base parameters~\cite{Khalil2004}, 
 $F_v \in \mathbb{R}^{n \times n}$ and $F(\cdot)  \in \mathbb{R}^n$  model viscous and nonlinear friction torques (i.e. $F_v$ is a positive definite matrix),
 and $\tau$ is the vector of 
 control inputs (i.e. desired actuators' torques) to be designed for achieving specific control objectives.  
 Furthermore, the following properties on the model~\eqref{eq:dynamics} hold true~\cite{Siciliano2008}:
 \begin{property}
  \label{hp:MisDP}
  The inertia matrix $M$ is a symmetric positive definite matrix, which implies
  \begin{IEEEeqnarray}{RCL}
	 \label{eq:MisDP}
	 \lambda_1(\pi) I_n  \leq M(q,\pi) \leq \lambda_2(\pi) I_n,\nonumber
  \end{IEEEeqnarray}
  with $\lambda_1$ and $\lambda_2$ two strictly positive constants.
 \end{property}
  \begin{property}
  \label{hp:SkwSym} 
  The matrix $\dot{M} - 2 C$ is skew-symmetric, i.e. 
  \begin{IEEEeqnarray}{RCL}
	 x^\top (\dot{M} - 2 C)x = 0, \quad \forall x \in \mathbb{R}^n.  \nonumber
  \end{IEEEeqnarray}
 \end{property}
\begin{property}
  \label{hp:Cbounded}
  The Coriolis matrix $C(q,\qD, \pi)$ satisfies 
  \begin{IEEEeqnarray}{RCL}
	 \label{eq:Cbounded}
	 |C(q,\qD,\pi)| \leq \lambda_0(\pi) |\qD|, \nonumber
  \end{IEEEeqnarray}
  for some bounded constant $\lambda_0$.
  \end{property}
\begin{property}
  \label{hp:Gbounded}
  The gravity vector $g(q, \pi)$ satisfies 
  \begin{IEEEeqnarray}{RCL}
	 \label{eq:Gbounded}
	 |g(q,\pi)| \leq \gamma_0(\pi) , \nonumber
  \end{IEEEeqnarray}
  for some bounded constant $\gamma_0$.
  \end{property}
  \begin{property}
  \label{hp:linearWRTparams}
  The model~\eqref{eq:dynamics} can be expressed linearly with respect to the system's base parameters $\pi$. Also, there exists
  a \emph{regressor matrix} $Y(\cdot) \in \mathbb{R}^{n \times p}$ such that  
  \begin{IEEEeqnarray}{RCL}
	 \label{eq:linearWRTparams}
	 M(q,\pi)\qDD &+& C(q,\qD,\pi)\xi + g(q,\pi) \nonumber \\  
	 &+& F_v(\pi)\xi + F(q,\qD,\pi) = Y(q,\qD,\xi,\qDD) \pi , \nonumber
  \end{IEEEeqnarray}
  for any vector $\xi \in \mathbb{R}^n$.
  \end{property}
  \noindent
  The matrix $Y(\cdot)$ is the so-called Slotine-Li regressor. As for an example, all above assumptions are satisfied in the case of rigid robot manipulators.
  
 \subsection{A known adaptive control result} 
 \label{subsec:slotine}
 Let $r(t)$ denote a time-varying reference trajectory for the joint variables $q$. Throughout the paper, we assume that: 
 \begin{assumption}
    \label{hp:reference}
    The reference trajectory $r(t)$ is bounded in norm on $\mathbb{R}^+$, and its first and second order derivatives are well-defined and bounded on this set.
 \end{assumption}
 We present below a revisited version of  an adaptive control scheme that ensures the asymptotic stabilization of the tracking error 
 \begin{IEEEeqnarray}{RCL}
    \label{eq:error}
    e := q -r
 \end{IEEEeqnarray}
 to zero without the knowledge of the inertial parameters $\pi$. The benefits of this new formulation will be clear in the next section. First, define: 
 \begin{IEEEeqnarray}{RCL}
    \label{eq:slotineVariables}
    s &:=& \qD - \xi,  \IEEEyessubnumber  \label{s} \\
    \tilde{\pi} &:=& \hat{\pi} - \pi, \IEEEyessubnumber
 \end{IEEEeqnarray}  
  where 
  $\tilde{\pi}$ is the inertial parameters estimation error. By considering 
  the dynamics $\dot{\hat{\pi}}$ and $\dot{\xi}$ as auxiliary control inputs, 
  fusing and reformulating the results \cite{Slotine1988} \cite{Spong1990} lead to the following lemma.
 \begin{lemma} 
 \label{th:slotine}
 Assume that Properties~\ref{hp:MisDP}-\ref{hp:linearWRTparams} and Assumption~\ref{hp:reference} hold true. Apply the following control laws to System~\eqref{eq:dynamics}
  \begin{IEEEeqnarray}{RCL}
    \label{eq:slotineControls}
    \tau &=& Y(q,\qD,\xi,\dot{\xi})\hat{\pi} - K s ,  \IEEEyessubnumber \label{tauSlotine}  \\ 
    \dot{\hat {\pi}} &=& -\Gamma Y^\top(q,\qD,\xi,\dot{\xi}) s, \IEEEyessubnumber \label{piHatSlotine} \\
    \dot{\xi} &=& \ddot{r} - \Lambda_1 \dot{e} - \Lambda_2 e ,  \IEEEyessubnumber \label{xiDotSlotine} 
 \end{IEEEeqnarray}
 with $K, \Lambda_1, \Lambda_2 \in \mathbb{R}^{n \times n}$ diagonal, constant positive definite matrices, and $\Gamma \in \mathbb{R}^{p \times p}$ 
 a constant positive definite matrix. Then, there exists a constant vector $\beta \in \mathbb{R}^n$ such that the equilibrium point 
 $(\int_0^t{e(s) \, ds},  e, s,\tilde{\pi}) = (\beta,0_n,0_n,0_p)$ of the 
 closed loop dynamics is globally stable, and the tracking error $e(t)$ converges to zero.
 \end{lemma}

 The proof is given in Appendix~\ref{proofSlotine}. The main difference between the above formulation and the one of \cite{Slotine1988} is that $\dot{\xi}$ is viewed as an auxiliary control input. Observe that the initial condition $\xi(0)$ can be arbitrary chosen thanks to the additional term $\Lambda_2 e$ in \eqref{xiDotSlotine}. This term plays the role of an integral action in the expression of \eqref{s}, and does not affect stability and convergence.
 For Lemma~\ref{th:slotine} to hold, it is assumed  that System~\eqref{eq:dynamics} is fully actuated.
 The following section proposes an extension of Lemma~\ref{th:slotine} to the case where System~\eqref{eq:dynamics} is underactuated.

\section{Main theoretical results}
\label{sec:main_result}

Assume that System~\eqref{eq:dynamics} endows only $m < n$ torque control inputs so that the first $k := n-m$ rows on the right hand side of Eq.~\eqref{eq:dynamics} are 
identically equal to zero, i.e.
 \begin{IEEEeqnarray}{RCL}
  \label{eq:tau}
  Y(q,\qD,\qD,\qDD) \pi = 
  \begin{pmatrix}
   0_k \\
   \bar{\tau}
  \end{pmatrix},
\end{IEEEeqnarray}
with $\bar{\tau} \in \mathbb{R}^m$. Now, partition the generalized coordinate vector~$q$ as follows
 \begin{IEEEeqnarray}{RCL}
  \label{eq:partitionq}
  q :=
  \begin{pmatrix}
    q_{n} \\
    q_c
  \end{pmatrix},
\end{IEEEeqnarray}
where $q_{n} \in \mathbb{R}^k$, $q_c \in \mathbb{R}^m$, and the suffixes ``$n$'' and ``$c$'' stand for \emph{noncollocated} 
and \emph{collocated}, respectively. 
Also, assume that the control objective is the asymptotic stabilization of the collocated joint coordinates $q_c$ about reference trajectories 
$r(t) \in \mathbb{R}^m$, i.e. the stabilization of the tracking error
\begin{IEEEeqnarray}{RCL}
    \label{eq:errorColloc}
    e := q_c -r
 \end{IEEEeqnarray}
 to zero. As before, we want to design control laws for this control objective without knowledge of the inertial parameters $\pi$.
 
Next theorem basically states that modulo a redefinition of the auxiliary control input $\dot{\xi}$, the laws~\eqref{eq:slotineControls} 
ensure the asymptotic stabilization of only the collocated joint variables.
 \begin{theorem} 
 \label{th:ExtensionSlotine}
 Assume that Properties~\ref{hp:MisDP}-\ref{hp:linearWRTparams} and Assumption~\ref{hp:reference} hold true. Partition the variables in~\eqref{s} and 
 the regressor $Y(\cdot)$ as follows: 
 \begin{IEEEeqnarray}{RCL}
    \label{partitioning}
    s := 
    \begin{pmatrix} 
	s_{\nc}  \\ 
	s_{\coll} 
    \end{pmatrix}, \quad
    \xi := 
    \begin{pmatrix} 
	\xi_{\nc}  \\ 
	\xi_{\coll} 
    \end{pmatrix}, \quad
    Y(\cdot) := 
    \begin{pmatrix} 
	Y_{\nc} (\cdot) \\ 
	Y_{\coll} (\cdot) 
    \end{pmatrix}, \quad  
\end{IEEEeqnarray}
where $s_{\nc},\xi_{\nc} \in \mathbb{R}^k $, $s_{\coll},\xi_{\coll} \in \mathbb{R}^{m} $, $Y_{\nc} \in \mathbb{R}^{k\times p} $, $Y_{\coll} \in \mathbb{R}^{m\times p} $.

Apply the following control laws to System~\eqref{eq:tau}
  \begin{IEEEeqnarray}{RCL}
    \label{eq:ExSlotine}
    \bar{\tau} &=& Y_{\coll}(q,\qD,\xi,\dot{\xi})\hat{\pi} - K s_{\coll} ,  \IEEEyessubnumber \label{tauExte} \\
    \dot{\hat {\pi}} &=& -\Gamma Y^\top (q,\qD,\xi,\dot{\xi} ) s, \label{paramUpsExt} \IEEEyessubnumber \label{piHatExte}  \\
    \dot{\xi} &{=}& 
    \begin{pmatrix}
       \dot{\xi}_{\nc} \\
       \dot{\xi}_{\coll}
    \end{pmatrix} {=} 
    \begin{pmatrix}
        \widehat{M}_{\nc}^{-1} \left[ K_{\nc} s_{\nc} {-} Y_{\nc}\left(q,\qD,\xi, \colvec {0_k}{\dot{\xi}_{\coll}}\right)\hat{\pi} \right]   \\
        \ddot{r} - \Lambda_1 \dot{e} - \Lambda_2 e
    \end{pmatrix}\IEEEeqnarraynumspace 
    \IEEEyessubnumber  \label{xiHatExte}
 \end{IEEEeqnarray}
 with $K, K_{\nc}, \Lambda_1, \Lambda_2 \in \mathbb{R}^{m \times m}$ diagonal, constant positive definite matrices, 
 and the matrix $\widehat{M}_{\nc}$ defined as the
 $k$\emph{th} order leading principal minor of the mass matrix $M$ evaluated with estimated base parameters, i.e. 
 \begin{IEEEeqnarray}{RCL}
    \label{nonactuatedMassEstim}
    \widehat{M}_{\nc} := S M(q,\hat{\pi}) S^\top   
 \end{IEEEeqnarray}
 where the selector $S$ is given by
  \begin{IEEEeqnarray}{RCL}
    \label{selector}
    S &: =& \begin{pmatrix} I_k &  0_{k \times m} \end{pmatrix}.  
 \end{IEEEeqnarray}
 

 Then, the following results hold.
\begin{enumerate}[i)]
  \item There exists a constant vector $\beta \in \mathbb{R}^m$ such that the equilibrium point $(\int_0^t{e(s) \, ds},  e, s,\tilde{\pi}) = (\beta,0_m,0_n,0_p)$ of the associated 
        closed loop dynamics is locally stable.
  \item Assume that the noncollocated joint velocities remain bounded, i.e. $\exists \delta > 0$ such that $|\dot{q}_{\nc}| < \delta \quad \forall t$. Then, 
  in addition to the local stability, the tracking error~$e(t)$ also converges to zero. 
\end{enumerate}
 \end{theorem}
 
 The proof is given in Appendix~\ref{proofExtSlotine}. 
 The interest of the invoked reformulation of classical adaptive schemes presented in Lemma~\ref{th:slotine} lies in the 
 similarity between the control laws~\eqref{eq:slotineControls} and~\eqref{eq:ExSlotine}. In particular, in both cases, the evolution of the variable $\xi$ can be obtained
 by numerical integration of its dynamics~$\dot{\xi}$. Also, when the system endows~$m$ unactuated degrees of freedom, it suffices to modify the first $m$ elements of this dynamics --~see Eq.~\eqref{xiHatExte}~-- to still ensure stability and convergence of the collocated joint coordinates.
 However, convergence is guaranteed when the noncollocated joint velocities $|\dot{q}_{\nc}|$ remain bounded. This requirement, which follows from the application of 
 Barbalat's Lemma, reflects physical limitations, mostly due to the energy exchanged between noncollocated and collocated joint variables. As a matter of fact,
 simulations we have performed suggest that friction effects play a role in guaranteeing the boundedness of $|\dot{q}_{\nc}|$, and consequently
 the convergence of the tracking error $e(t)$ to zero.
 
 The local nature of the controls~\eqref{eq:ExSlotine} is due to the fact that the matrix~\eqref{nonactuatedMassEstim} may not be invertible far from 
 the point $\tilde{\pi}=0_p$. Observe that the invertibility of~\eqref{nonactuatedMassEstim} in a neighborhood of this point is guaranteed by Property~\ref{hp:MisDP}, which
 implies that each leading principal minor of the mass matrix $M(q,\pi)$ is positive definite, and therefore invertible. 

 The non-invertibility of the matrix~\eqref{nonactuatedMassEstim} is related to the 
 \emph{standard parameters}\footnote{For example, the standard parameters of a rigid body consists in a ten-dimensional vector composed of the mass, 
 the three first moments of mass, and the six elements of the inertia matrix~\cite{Khalil2004}.} associated with the estimated base parameters $\hat{\pi}$. 
 In particular, when the \emph{standard parameters} associated with an estimate $\hat{\pi}$ are not physical consistent 
 (e.g. a negative mass of a rigid body composing the underlying mechanical system) the inertia matrix $M(q,\hat{\pi})$ 
 may not be positive definite~\cite{Yoshida2000}. This problem would  be avoided if the adaptation law $\dot{\hat{\pi}}$ guaranteed an evolution $\hat{\pi}(t)$ such that 
 the associated standard parameters were always physical consistent. Such an adaptation law is under preparation and will be presented and discussed in a forthcoming paper. 
 
 Now, let us remark that if 
  \begin{IEEEeqnarray}{RCL}
    \det{\left(  \widehat{M}_{\nc}(q(t),\hat{\pi}(t)) \right) } > 0 \quad \forall t
    \label{detGreaterZero}
 \end{IEEEeqnarray}
 independently of the initial conditions, the laws~\eqref{eq:ExSlotine} ensure global stability. 
 However, this is not always the case.  To avoid a possible ill-conditioning of the laws~\eqref{eq:ExSlotine}, 
 a desingularization policy must be defined when the above determinant gets close to zero. 
 We present below the policy used in this paper.
 
 Observe that the time derivative of \eqref{detGreaterZero} depends upon the adaptation law $\dot{\hat{\pi}}$. Consequently, it is theoretically possible to modify the law~\eqref{piHatExte} in order to ensure that 
 the determinant of $\widehat{M}_{\nc}$ never decreases below a certain threshold. Next Lemma presents such a modification of the adaptation law $\dot{\hat{\pi}}$.
  \begin{lemma}
    \label{lemma:desingularization}
    Consider the laws~\eqref{eq:ExSlotine} with  the adaptation law redefined as follows
    \begin{IEEEeqnarray}{RCL}
    	\dot{\hat {\pi}} &=& -\Gamma \left[ Y^\top(q,\qD,\xi,\dot{\xi}) s - \eta \delta \right] , \label{paramUpsExtDes} 
    \end{IEEEeqnarray}
    with $\eta \in \mathbb{R}$ and $\delta \in \mathbb{R}^p$ given by:
    \begin{IEEEeqnarray}{RCL}
	\label{etaDelta}
	\vspace*{0.2cm}
	\eta &{:=}& 
	  \left\{
	    \begin{array}{l l}
	      \hspace{-0.2cm} 0 &  \hspace{-0.55cm}	\text{if } \tr \left(\widehat{M}_{\nc}^{-1} \Upsilon \right) {\geq} 0 \text{ or }  \det{ \left(\widehat{M}_{\nc} \right) } {>}  \varepsilon    \\
	      \hspace{-0.2cm} {-}\frac{\tr(\widehat{M}_{\nc}^{-1} \Upsilon)}{\delta^T \Gamma \delta} &     \text{otherwise}
	    \end{array} \right.  \IEEEeqnarraynumspace \IEEEyessubnumber \label{def:eta}  \vspace{0.1cm} \\
    	\delta &{:=}& \sum_{i = 1}^{k}{Y_{M_{\nc}}^\top(q, e_i) \widehat{M}_{\nc}^{-1} e_i}, \IEEEyessubnumber \label{delta} 
    \end{IEEEeqnarray}
    where $\varepsilon \in \mathbb{R}^+$,
    \begin{IEEEeqnarray}{RCL}
    	Y_{M_{\nc}} &:=& S \left[ Y\left(q,0_n,0_n, \colvec{e_i}{0_m} \right) - Y(q,0_n,0_n,0_n)\right],\IEEEeqnarraynumspace  \IEEEyessubnumber \\
	\Upsilon &:=& \left( \upsilon_1, \dots , \upsilon_i, \dots , \upsilon_k  \right), \IEEEeqnarraynumspace    \IEEEyessubnumber \label{def:upsilon}\\
	\upsilon_i &:=& \frac{\partial}{\partial q} \Big[ Y_{M_{\nc}}\hat{\pi} \Big] \qD -Y_{M_{\nc}}\Gamma Y^\top(q,\qD,\xi,\dot{\xi})s , \IEEEyessubnumber \IEEEeqnarraynumspace
    \end{IEEEeqnarray}
    and 
    $e_i \in \mathbb{R}^k$ denotes a vector of~$k$ zeros 
    except for the $i$\emph{th} coordinate, which is equal to one. 
    
    Then, the following results hold.
    \begin{enumerate}[i)]
	\item If $\det{\left(  \widehat{M}_{\nc} \right) } > 0$, then $|\delta| > 0$.
	\item Assume that $\det{\left(  \widehat{M}_{\nc} \right) }(0) > \varepsilon$. Then, \[\det{\left(  \widehat{M}_{\nc} \right) }(t) \geq \varepsilon \quad \forall t.\]
\end{enumerate}
 \end{lemma}
 
 \begin{figure}[t!]
    \begin{center}
        \includegraphics[width=.9\columnwidth]{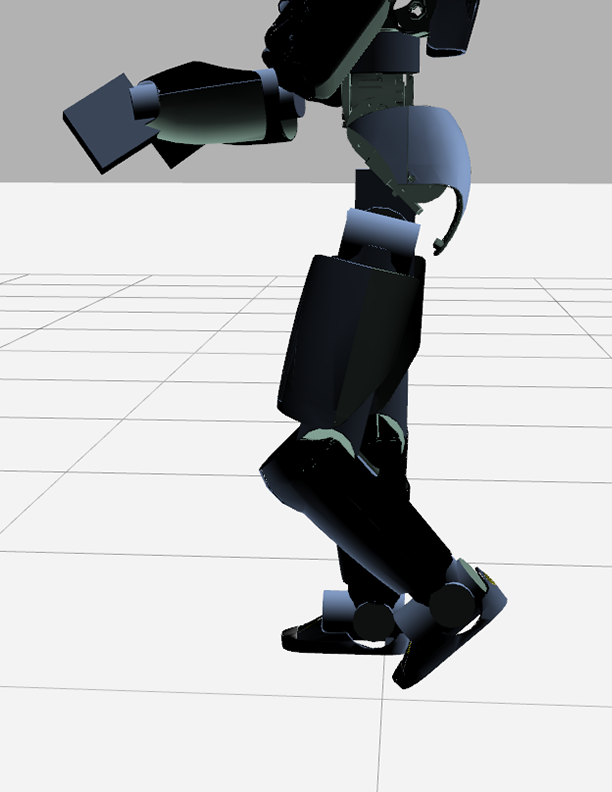}
        \caption{iCub model in the Gazebo environment.}
        \label{fig:icub_gazebo}
    \end{center}
\end{figure}
 
The proof is in Appendix~\ref{proofDesingula}. This Lemma states that it is always possible to maintain the determinant of $\widehat{M}_{\nc}$ above a certain threshold $\varepsilon$.
In fact, the desingularizing term~\eqref{etaDelta} would be ill-conditioned only at $|\delta| = 0$, but this never occurs when $\det{(  \widehat{M}_{\nc} ) } > 0$ -- see the result $i)$.

Clearly, the larger the threshold $\varepsilon$, the larger the influence of the desingularizing term $\eta \delta$ on the stability result of Theorem~\ref{th:ExtensionSlotine}.
Consequently, this threshold must be tuned depending on the specific application, which defines the value of $\det{(  \widehat{M}_{\nc} ) }$ at $\hat{\pi} = \pi$.
Simulations and experimental results presented next show that the influence of this desingularizing term does not significantly affect the practical stability and boundedness
of the tracking error $e(t)$.

%
%

\section{Simulations and Experimental Results}
\label{sec:simulation}

\begin{figure}[t!]
    \begin{center}
        \includegraphics[width=.9\columnwidth]{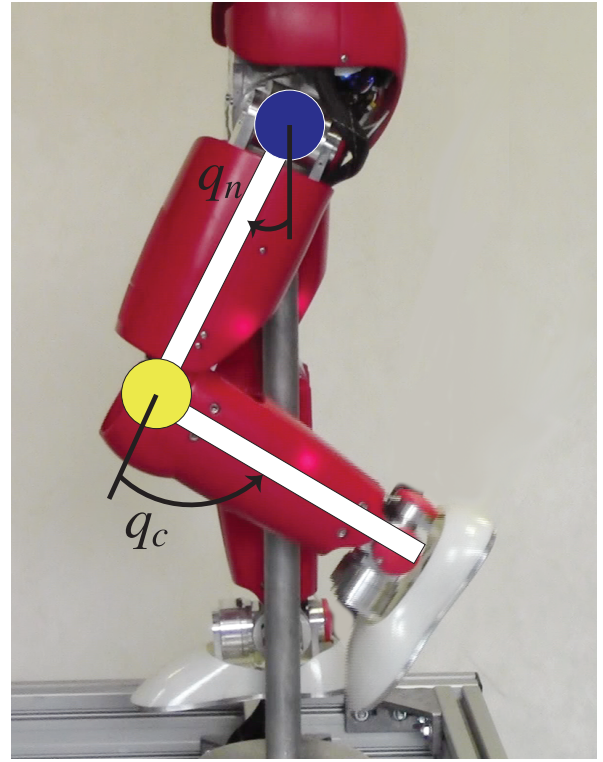}
        \caption{Experimental setup. The two-link pendulum is depicted in overlay. Blue circle: underactuated joint ($q_n$); yellow circle: actuated joint ($q_c$).}
        \label{fig:icub-system}
    \end{center}
\end{figure}

In this section, we test the control laws~\eqref{tauExte}-\eqref{xiHatExte}-\eqref{paramUpsExtDes} 
first through simulations, and then through experiments carried out on a two-link manipulator with rotational joints. 
In particular, the robot consists in the \emph{hip} (nonactuated joint) and the \emph{knee} (actuated joint) of the iCub humanoid robot, while the robot's ankle is kept fixed  
(see Figure~\ref{fig:icub-system}). The underactuated system is then simulated in the ``Gazebo''\footnote{Gazebo is the official simulator of the Darpa 
Robotic Challenge, and is developed by the Open Source Robotics Foundation (OSRF). The official website is \url{http://www.gazebosim.org} } environment. 
Furthermore, the YARP middle-ware interfaces the robot, either real or simulated, to our controller 
implementation\footnote{The implementation is available at the \href{https://github.com/robotology/codyco}{CoDyCo repository}}. 
The main aim of the simulation campaign is to determine a set of gains to use during the experiments.



\begin{figure*}[t!]
    \centering
    \begin{subfigure}[t]{\columnwidth}
        \includegraphics[width=\columnwidth]{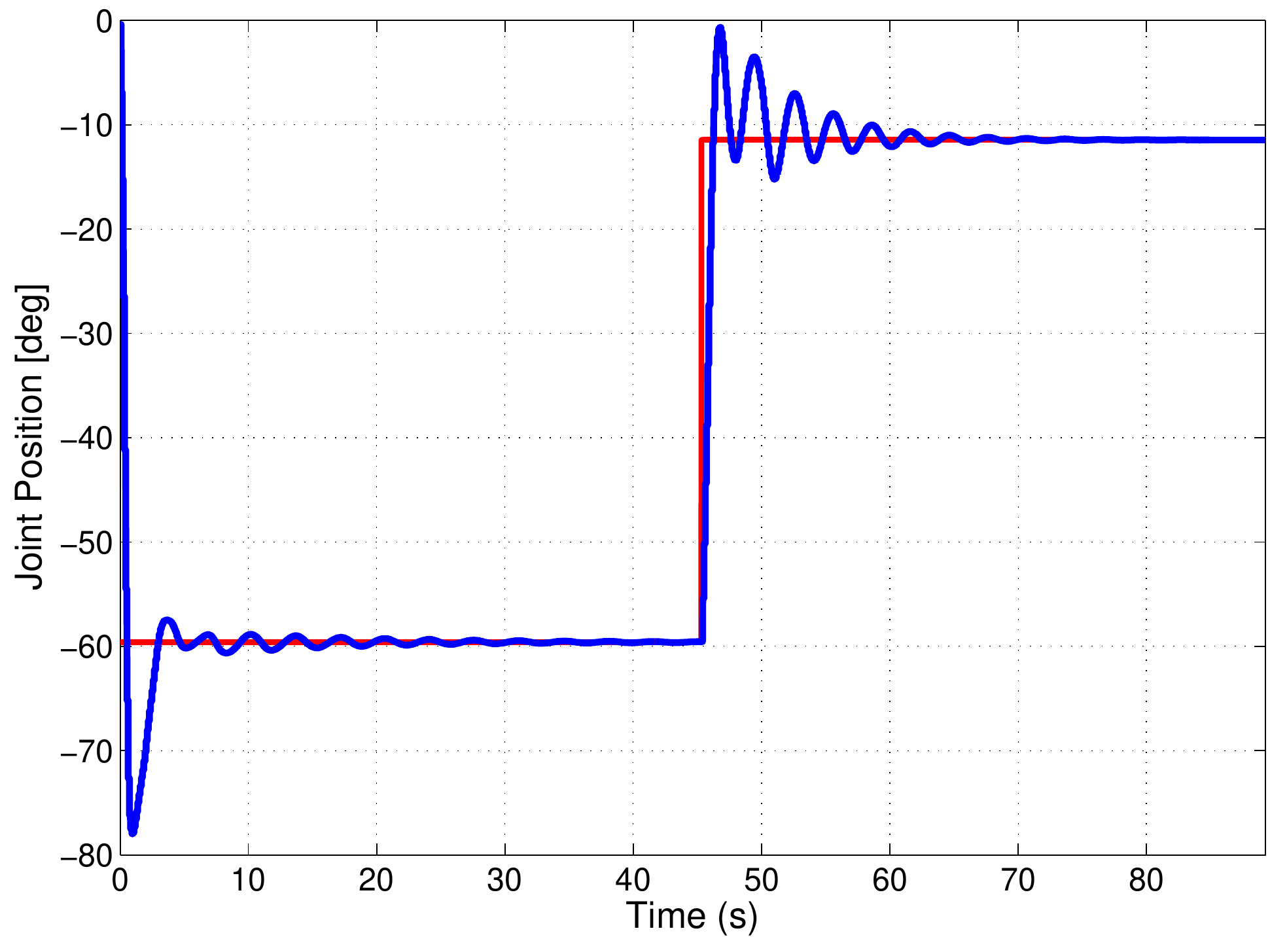}
        \caption{Knee angle $q_\coll$ (blue) and its reference value (red).}
        \label{fig:sim_constant_position}
    \end{subfigure}
    ~
    \begin{subfigure}[t]{\columnwidth}
        \includegraphics[width=\columnwidth]{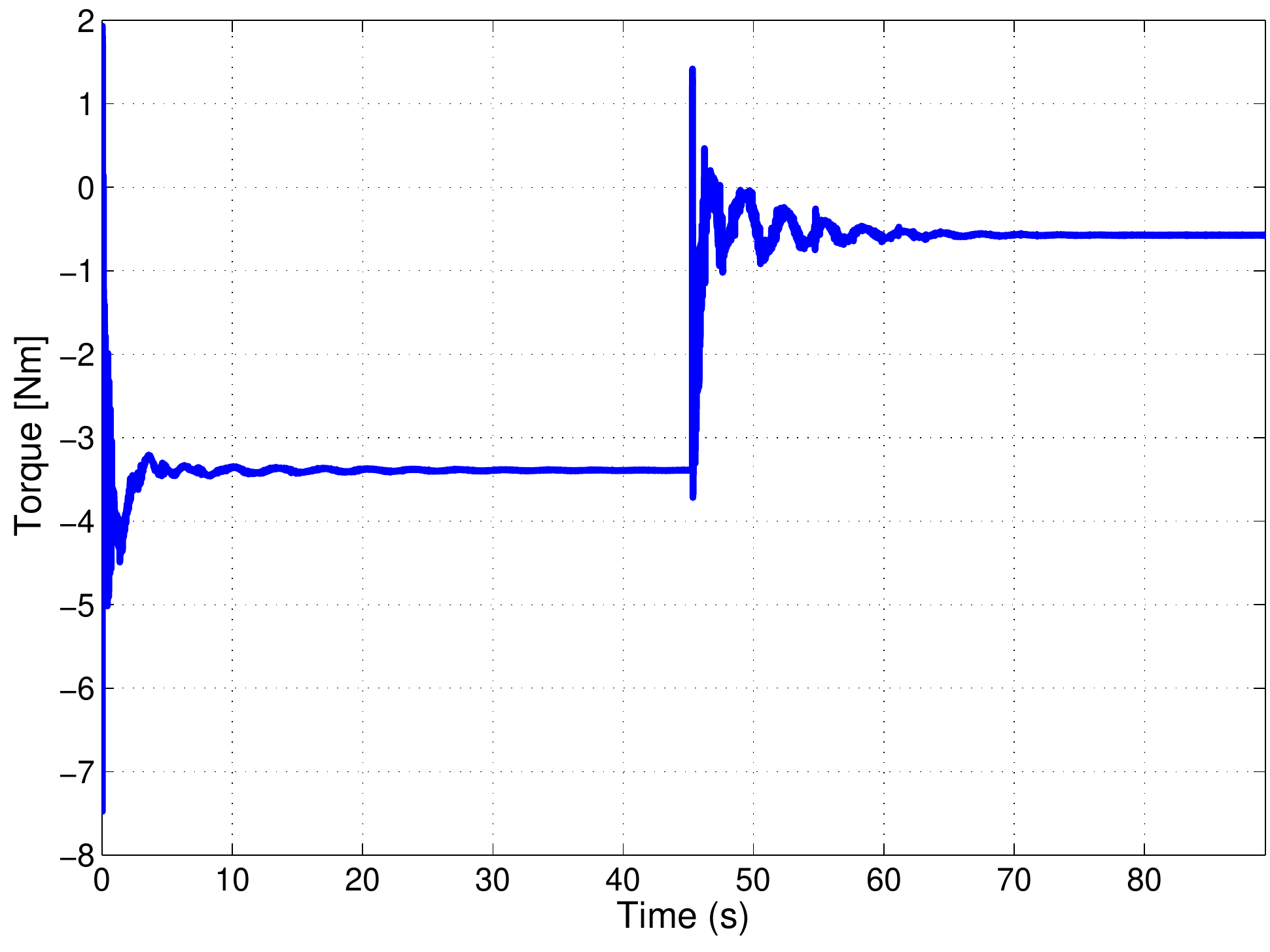}
        \caption{Desired torque for the actuated joint.}
        \label{fig:sim_sin_torque}
    \end{subfigure}
    \caption{Simulation results.}
    \label{fig:sim_plot}
\end{figure*}

The laws~\eqref{tauExte}-\eqref{xiHatExte}-\eqref{paramUpsExtDes} require to compute the regressor $Y(\cdot)$ of a two-link manipulator~\cite[p. 149]{Siciliano2009}. 
This regressor is computed with only viscous friction terms, i.e. $F(\cdot) \equiv 0$. 
The main reasons why are the following ones.
\begin{enumerate}[i)]
  \item The simulator ``Gazebo'' currently supports only viscous friction. 
  \item The iCub humanoid robot is equipped with a low-level torque control loop that is in charge of stabilizing \emph{any desired joint torque}~\cite{Fumagalli2010, Fumagalli2012}:
	this loop is supposed to compensate for friction effects, but this compensation is never perfect. 
	Then, the terms of viscous friction left in the regressor account for imperfect friction compensations during adaptive control.
\end{enumerate}


Figure~\ref{fig:sim_plot} depicts typical simulation results obtained with the laws~\eqref{tauExte}-\eqref{xiHatExte}-\eqref{paramUpsExtDes} 
evaluated with
\begin{itemize}
    \item $\Lambda_1 = 5$,
    \item $\Lambda_2 = 1$,
    \item $K = 0.1 I_2$,
    \item $\Gamma = 0.1 I_8$,
    \item $\varepsilon = 5$,
    \item $\hat{\pi}(0) = \left(1.5 \text{, }  {-}0.11 \text{, } 0.01 \text{, } 2\text{, } {-}0.24 \text{, } 0.08 \text{, } 0.05\text{, } 0.05 \right)$,
    \item $\xi(0) = 0_2 \text{, } \beta = 0$.
\end{itemize}
The desired trajectory for the knee's joint was piecewise constant, and more precisly equal to: $r = -60^\circ$,$0 \leq t \leq 46s$, $r = -12^\circ$, $46s \leq t \leq 90s$. 
Figure~\ref{fig:sim_plot} shows that convergence of the tracking error is quickly achieved, while the actuation torque remains within the physical limits. 
By using the above gains, we went one step further and applied the laws~\eqref{tauExte}-\eqref{xiHatExte}-\eqref{paramUpsExtDes} to the aforementioned two-link robot.


\begin{figure*}
    \centering
    \begin{subfigure}[t]{\columnwidth}
        \includegraphics[width=\columnwidth]{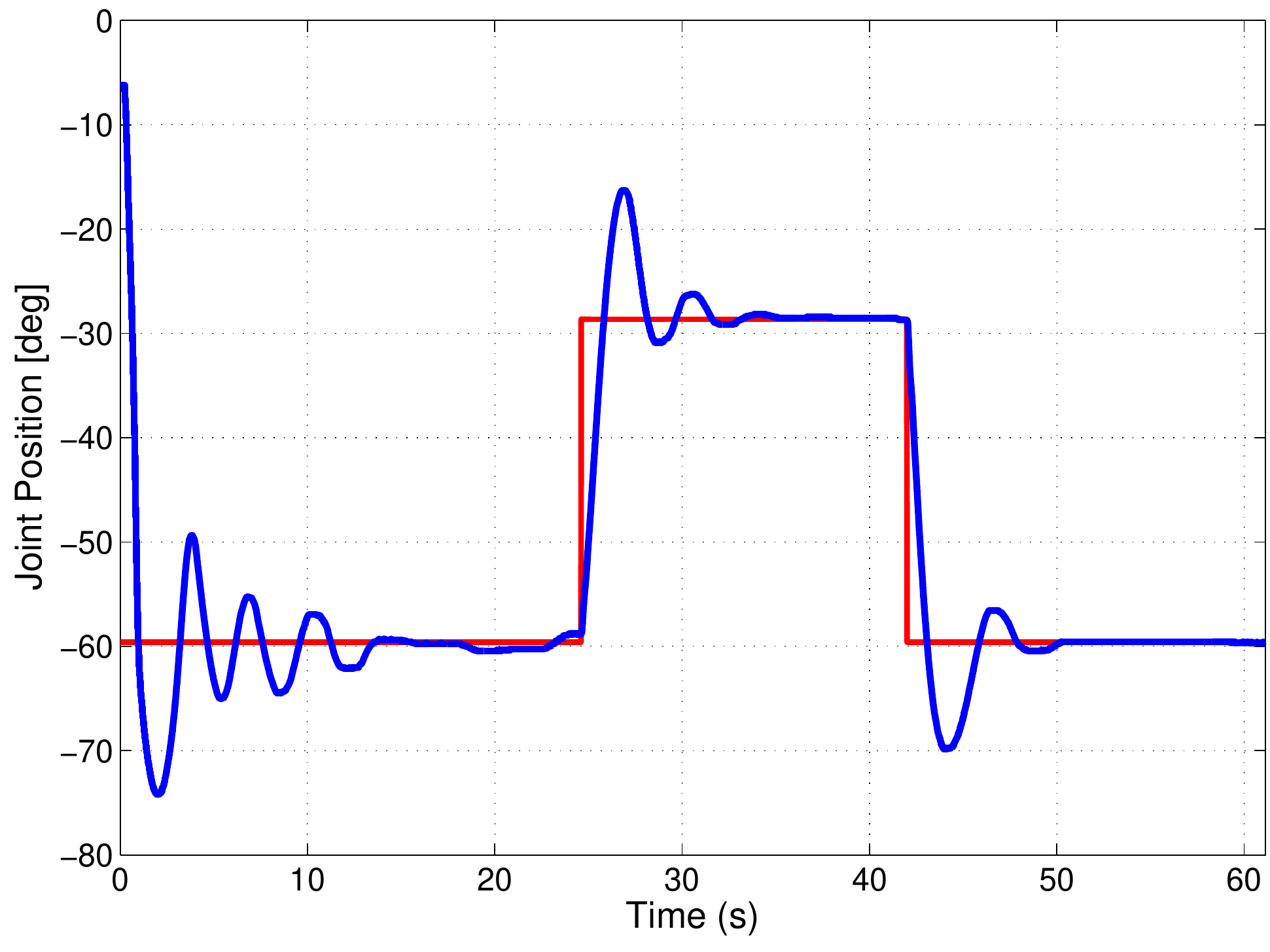}
        \caption{Knee angle (blue) and its reference value (red) (as Equation~\ref{ref1}).}
        \label{fig:robot_constant_position}
    \end{subfigure}
    ~
    \begin{subfigure}[t]{\columnwidth}
        \includegraphics[width=\columnwidth]{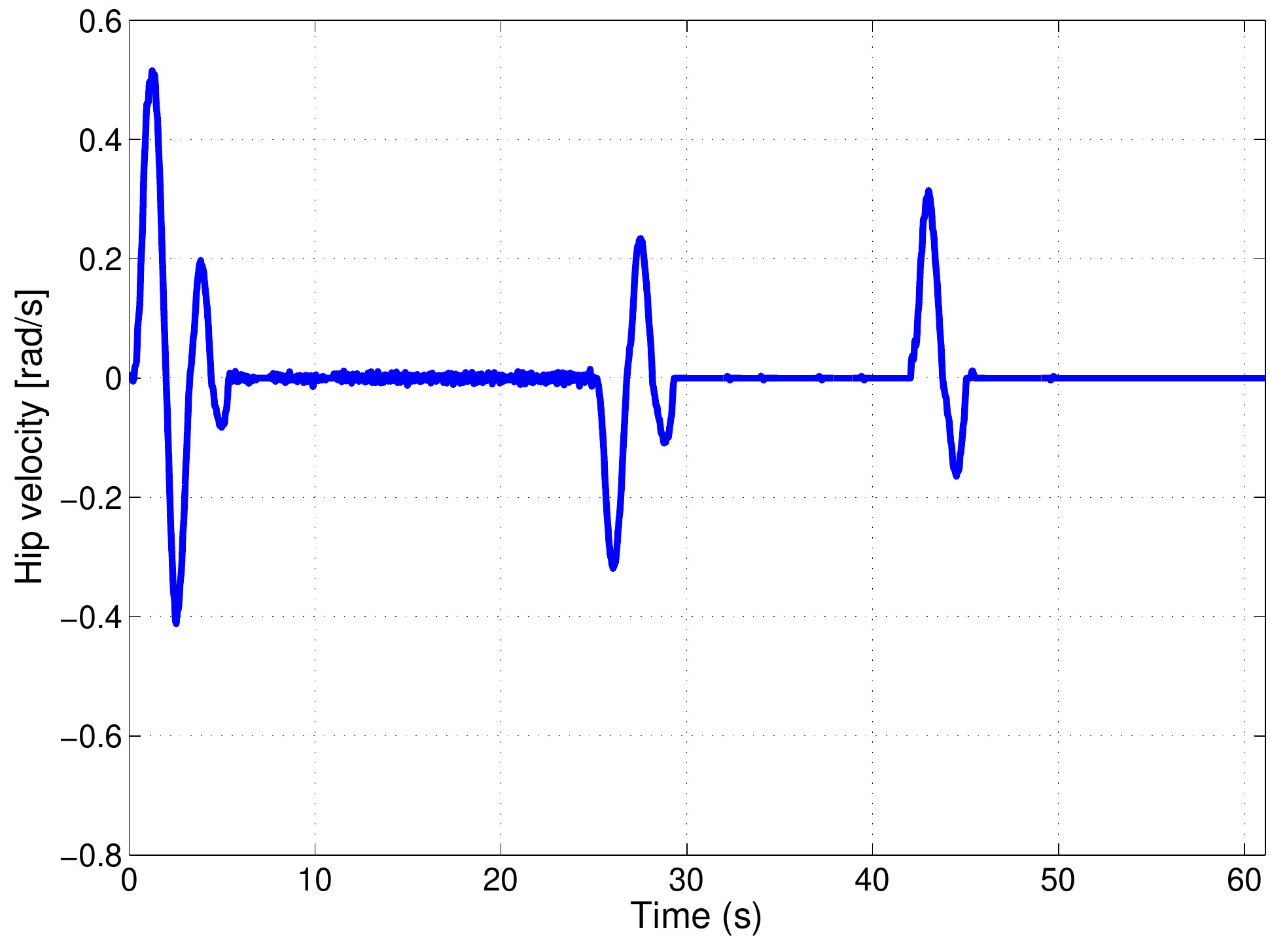}
        \caption{Velocity of the hip $\dot{q}_\nc$.}
        \label{fig:robot_const_dq1}
    \end{subfigure}
    
    \begin{subfigure}[t]{\columnwidth}
        \includegraphics[width=\columnwidth]{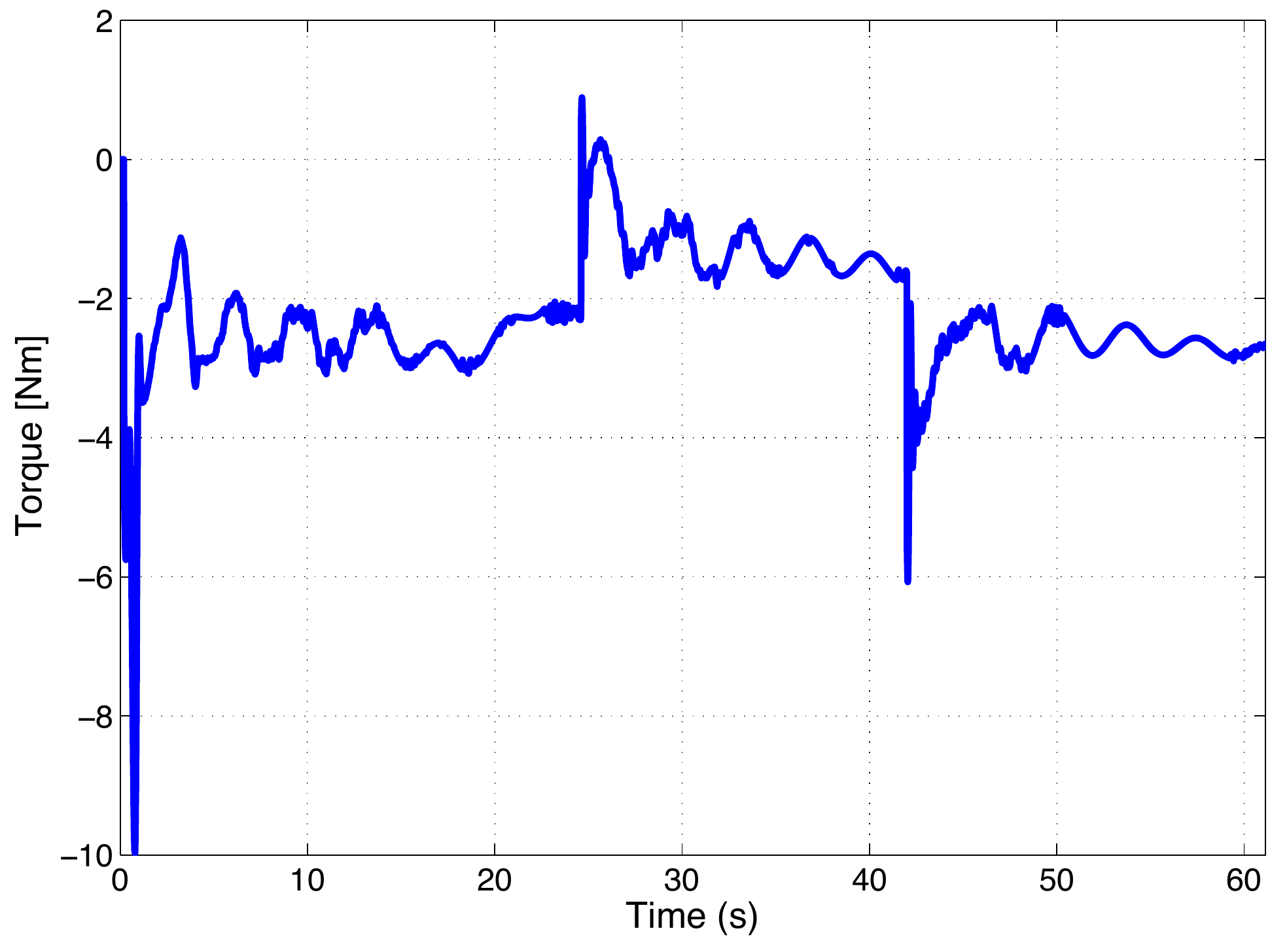}
        \caption{Desired knee joint torque to be stabilized by the low-level torque control loop.}
        \label{fig:robot_constant_torque}
    \end{subfigure}
    ~
    \begin{subfigure}[t]{\columnwidth}
        \includegraphics[width=\columnwidth]{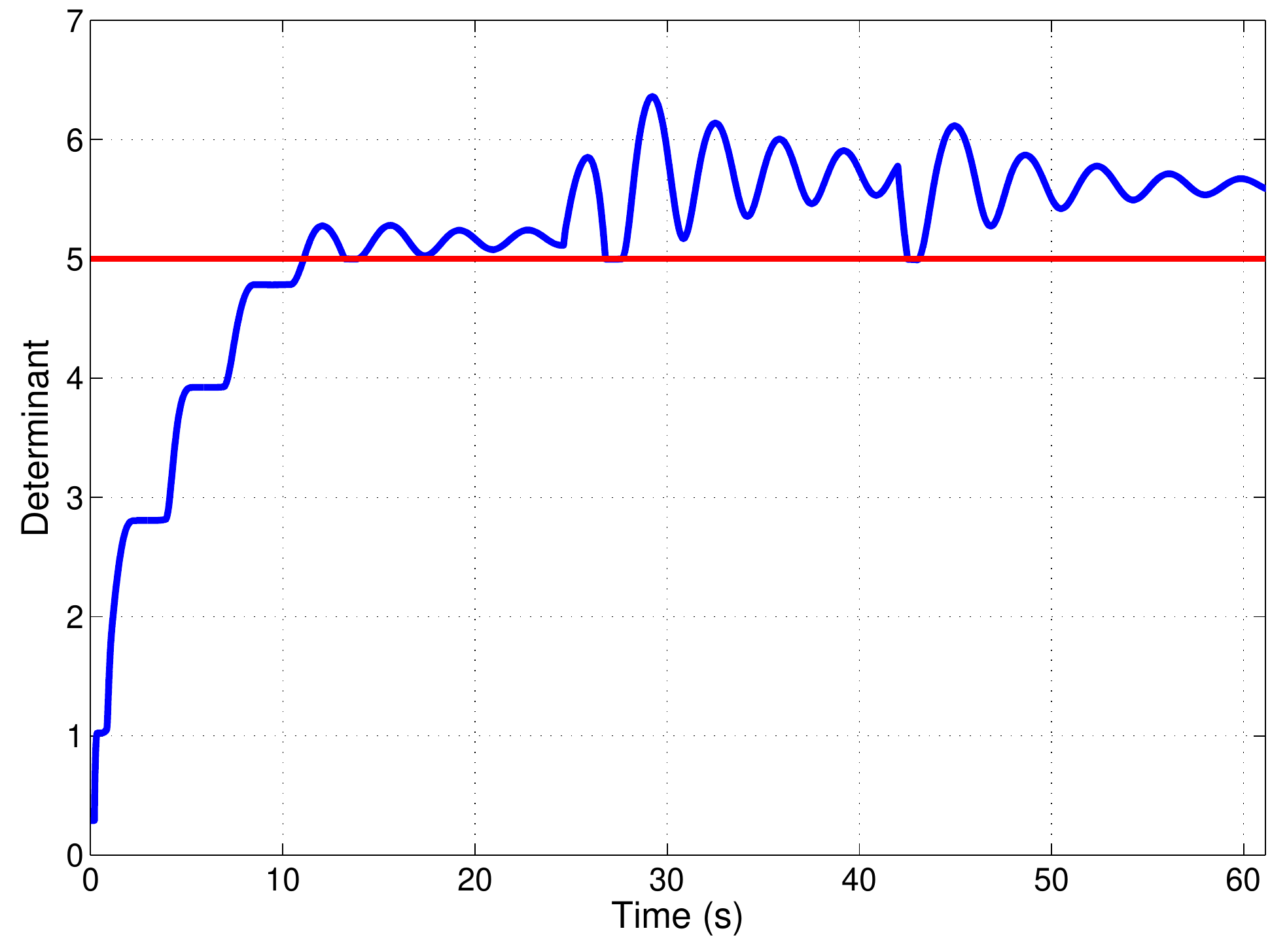}
        \caption{Determinant of the matrix $\widehat{M}_n$ (blue) and threshold $\varepsilon = 5$ (red).}
        \label{fig:robot_constant_determinant}
    \end{subfigure}    
        
    \caption{Experimental results for constant reference value \eqref{ref1}.}
    \label{fig:robot_constant}
\end{figure*}

Two experiments with two different reference trajectories $r(t)$ were conducted on the real robot. They are:

\begin{inparaenum}[i)]
    \item a piecewise constant reference defined by 
\begin{IEEEeqnarray}{RCL}
	\label{ref1}
	r(t) =
	  \left\{
	    \begin{array}{l l}
	      -60^\circ  \hspace{1.2cm} 0 \leq t < 25, \\
		-32^\circ \hspace{1cm} 25 \leq t < 43, \\
		-60^\circ  \hspace{1cm} 43 \leq t < 62; \\
	    \end{array} \right. 
\end{IEEEeqnarray} 
    \item a time varying reference $r(t)$ defined by (in degrees):
    \begin{IEEEeqnarray}{RCL}
	\label{ref2}
	r(t) =
	  \left\{
	    \begin{array}{l l}
	      -60 + 35 \sin(2 \pi 0.2 t) \quad  \hspace{0.7cm} 0 \leq t < 53, \\
	      -60 + 35 \sin(2 \pi 0.3 t) \quad  \hspace{0.55cm} 53 \leq t < 90, \\
          -60 + 35 \sin(2 \pi 0.35 t)\quad  \hspace{0.4cm} 90 \leq t < 110. \\
	    \end{array} \right. \IEEEeqnarraynumspace
\end{IEEEeqnarray} 
\end{inparaenum}

Figure~\ref{fig:robot_constant} depicts the results of the experiment i). Observe that the tracking error converges to zero with a settling time approximately
equal to that of the simulation result (compare Figures~\ref{fig:sim_plot}a and~\ref{fig:robot_constant_position}). However, unmodeled friction effects and imperfect 
tracking of the low-level torque control loop reflect in reduced overshoots and zero hip velocity close to zero tracking error (see Figure~\ref{fig:robot_const_dq1}). 
Note also that the initial conditions
for $\hat{\pi}$ render the determinant of $\widehat{M}_n$ at $t = 0$ smaller than the chosen threshold $\varepsilon~=~5$ (see Figure~\ref{fig:robot_constant_determinant}).
Then, this determinant can only increase because of the desingularization of Lemma~\ref{lemma:desingularization}, and once it goes beyond the threshold, it never decreases 
$\varepsilon~=~5$. It is important to observe that this experiment is basically quasi-static, so the coupling effects between the collocated and noncollocated joints
are not preponderant.

\begin{figure*}
    \centering
    \begin{subfigure}[t]{\columnwidth}
        \includegraphics[width=\columnwidth]{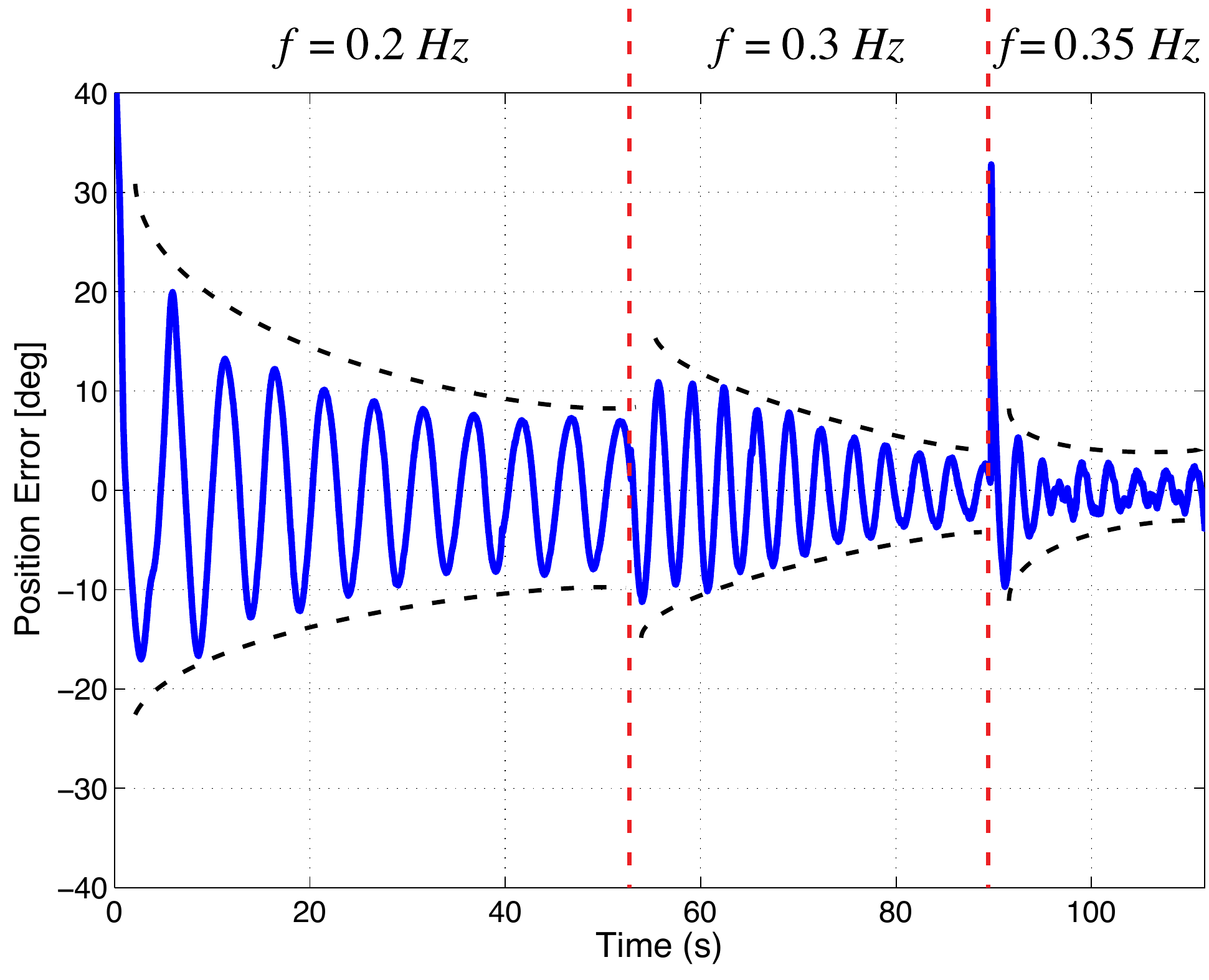}
        \caption{Knee angle (blue) and its reference value (red) (as Equation~\ref{ref2}).}
        \label{fig:robot_sin_error}
    \end{subfigure}
    ~
    \begin{subfigure}[t]{\columnwidth}
        \includegraphics[width=\columnwidth]{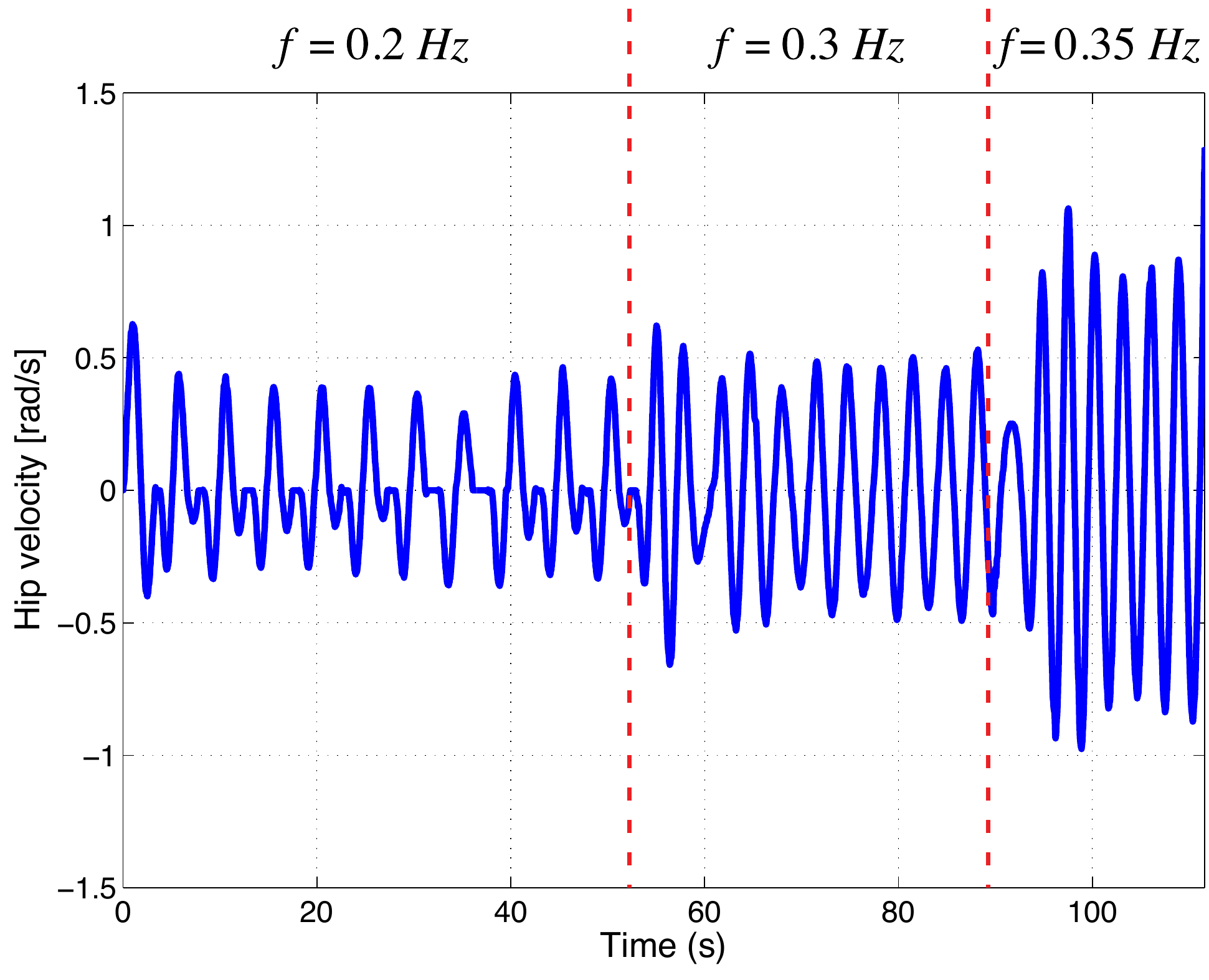}
        \caption{Velocity of the hip $\dot{q}_\nc$.}
        \label{fig:robot_sin_dq1}
    \end{subfigure}
    
    \begin{subfigure}[t]{\columnwidth}
        \includegraphics[width=\columnwidth]{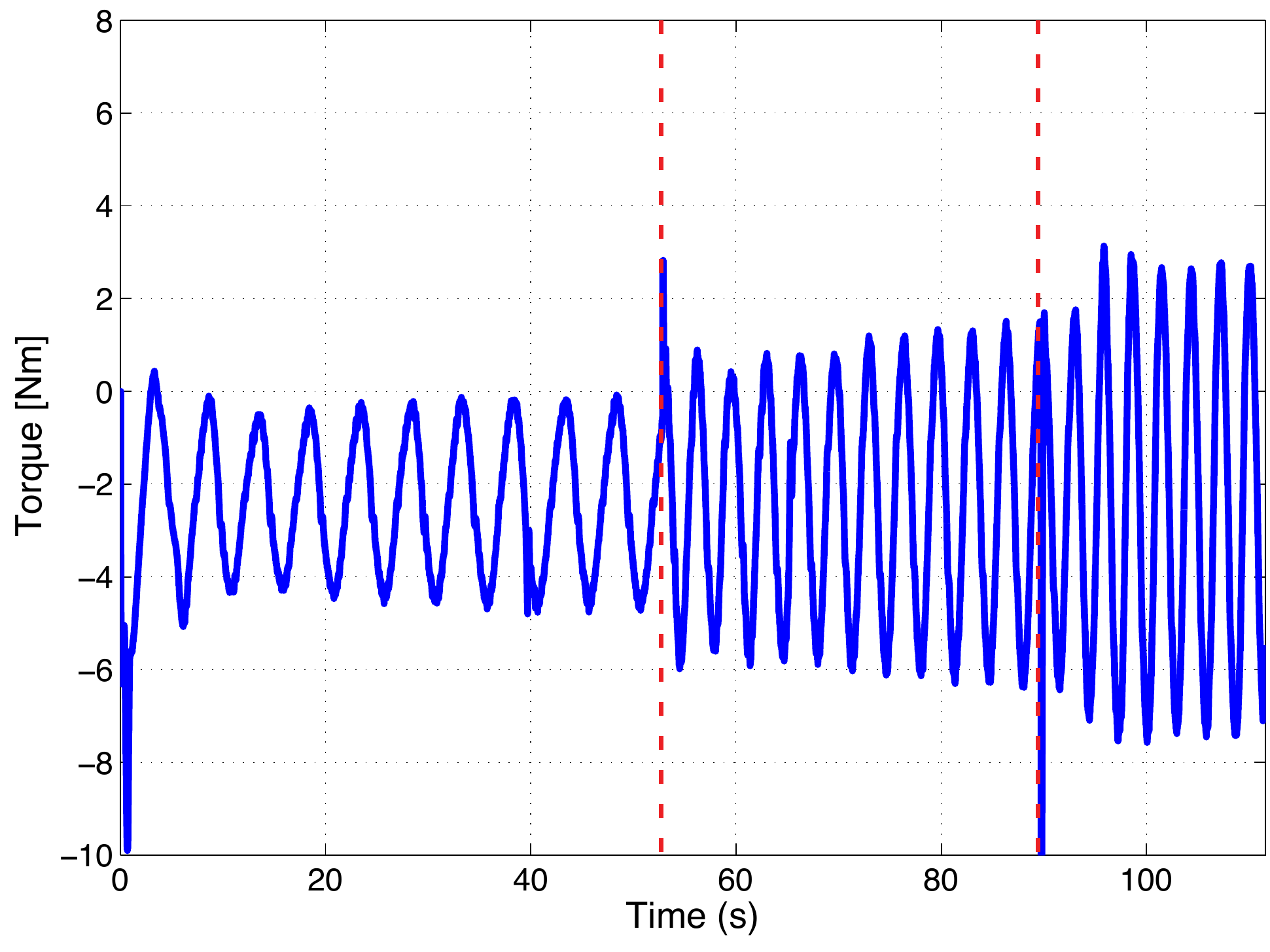}
        \caption{Desired knee joint torque to be stabilized by the low-level torque control loop.}
        \label{fig:robot_sin_torque}
    \end{subfigure}
    ~
    \begin{subfigure}[t]{\columnwidth}
        \includegraphics[width=\columnwidth]{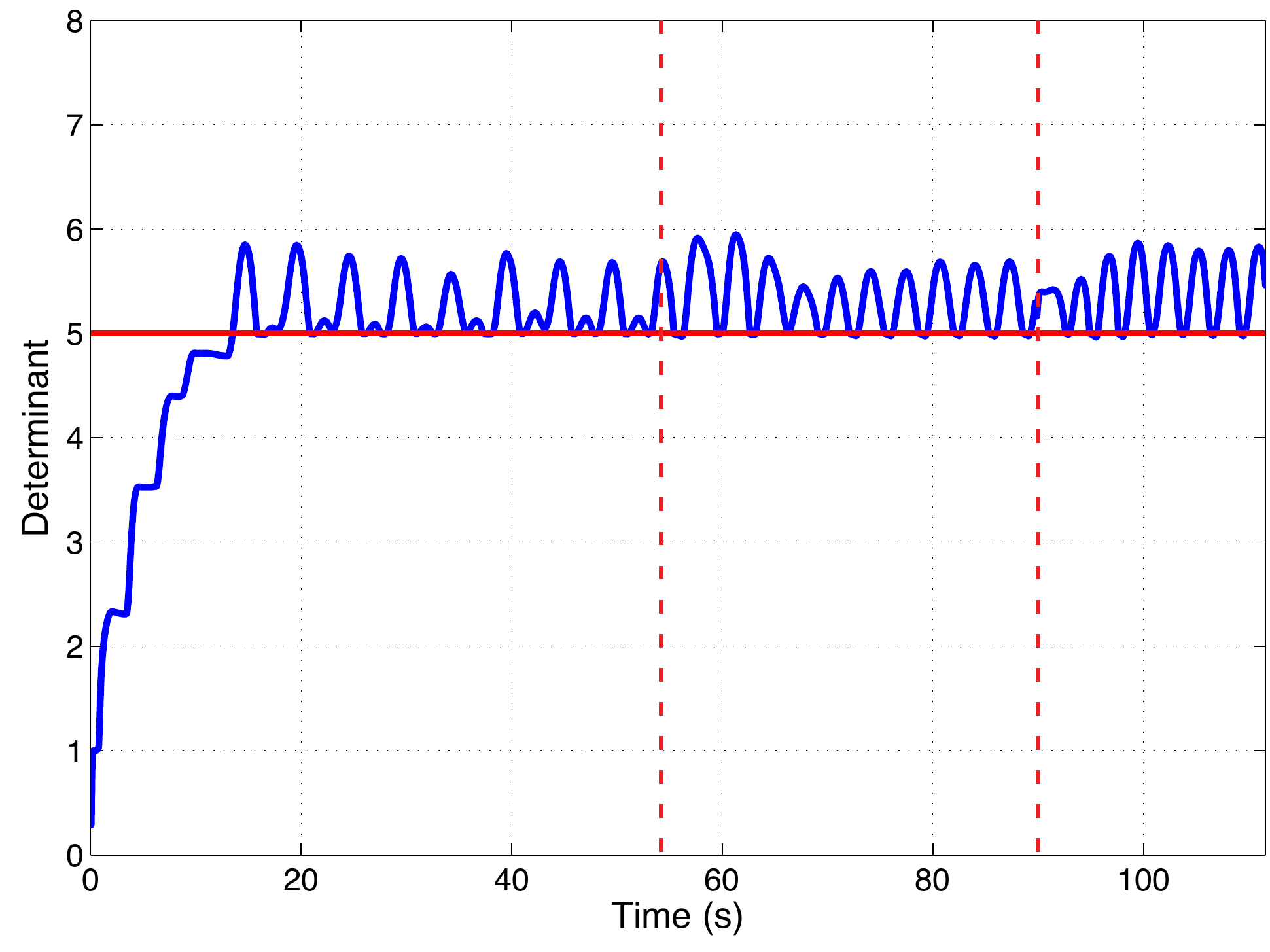}
        \caption{Determinant of the matrix $\widehat{M}_n$ (blue) and threshold $\varepsilon = 5$ (red).}
        \label{fig:robot_sin_determinant}
    \end{subfigure}
    
    \caption{Experimental results for time-varying reference trajectory \eqref{ref2}.}
    \label{fig:robot_sin}
\end{figure*}

Figure~\ref{fig:robot_sin} shows the experimental results obtained by applying the aforementioned laws with the time-varying reference trajectory 
given by~\eqref{ref2}. Observe that despite rapid variations of the \emph{desired} control torque (see Figure~\ref{fig:robot_sin_torque}), 
the tracking error remains bounded and slowly converges to zero. Also, note that coupling effects between the hip and knee joints are no longer negligible,
since the hip velocity achieves peaks of about $60 \ [deg/s]$ (see Figure~\ref{fig:robot_sin_dq1}).
These rapid variations of the joint coordinates $q$ cause large oscillations of the determinant of the matrix $\widehat{M}_n(q,\hat{\pi})$ (see Figure~\ref{fig:robot_sin_determinant}).
However, the desingularization of Lemma~\ref{lemma:desingularization} forces the determinant to remain above the threshold $\varepsilon = 5$.


\section{Conclusion} 
\label{sec:conclusion}

We presented an extension of the adaptive control method~~\cite{Slotine1988}, which was developed for fully actuated manipulators, 
to the case of underactuated mechanical systems.
Local stability and convergence of the collocated variables were demonstrated by using Lyapunov and Barbalat arguments.
Compared to existing results, our approach does not make use of any  acceleration measurements, 
thus avoiding altogether causality concerns.
The control results were validated with both simulations performed in the Gazebo environment, and with an implementation 
on a two-link manipulator obtained from the iCub humanoid robot.
It was beyond the scope of this work to address the classical, and well known, drawbacks of adaptive control schemes~\cite{Anderson2005a}.

%

Although the control laws presented in this paper are continuous, the associated basin of attraction of the equilibrium point is local.
This local nature is due to the fact that the presented control laws rely on the invertibility of the system's inertia matrix along the 
estimated system's model. This matrix may not be invertible for physical inconsistent \emph{base parameters}~\cite{Yoshida2000}. Then, our goal is to design
an estimation dynamics such that the associated base parameters are always physical consistent. In this case,
the control laws presented in this paper would guarantee global stability and convergence.


\appendix
\setcounter{subsection}{0}

\subsection{Proof of Lemma~\ref{th:slotine}}
\label{proofSlotine}

First, from Eq.~\eqref{xiDotSlotine} observe that the variable $\xi(t)$ can be obtained by integration, i.e.
\begin{IEEEeqnarray}{RCL}
  \xi(t) = \dot{r} - \Lambda_1 e - \Lambda_2 y, \nonumber
\end{IEEEeqnarray}
with $y := \int_0^t{e(s) \, ds} - \beta$, and $\beta \in \mathbb{R}^n$ a properly chosen constant. 
Now, consider the following
candidate Lyapunov function
\begin{IEEEeqnarray}{RCL}
  \label{lyapunovSlotine}
  V{:=} \frac{1}{2}\left[ s^\top M s {+} \tilde{\pi}^\top \Gamma^{-1}\tilde{\pi} {+} 2 e^\top K \Lambda_1 e + 2 y^\top \Lambda_1 K \Lambda_2 y\right]. 
  \IEEEeqnarraynumspace
\end{IEEEeqnarray}
Note that since $K, \Lambda_1,$ and $\Lambda_2$ are diagonal matrices, then the products $K \Lambda_1$ and $\Lambda_1 K \Lambda_2 $ are diagonal and 
positive definite matrices. In view of Properties~\ref{hp:SkwSym},~\ref{hp:linearWRTparams} and the controls~\eqref{tauSlotine}-\eqref{piHatSlotine},
one easily verifies that the time derivative of~\eqref{lyapunovSlotine} yields,
\begin{IEEEeqnarray}{RCL}
  \label{lyapunovDerivativeSlotine}
  \dot{V}{=} -s^\top K s -s^\top F_v s  + 2 e^\top K \Lambda_1 \dot{e} + 2 y^\top \Lambda_1 K \Lambda_2 e. 
  \IEEEeqnarraynumspace
\end{IEEEeqnarray}
Recall that $F_v$ is a positive definite matrix. By substituting
\begin{IEEEeqnarray}{RCL}
  \label{sXi}
  s = \qD - \xi = \dot{e} + \Lambda_1 e + \Lambda_2 y \nonumber
  \IEEEeqnarraynumspace
\end{IEEEeqnarray}
in the first term on the right hand side of~\eqref{lyapunovDerivativeSlotine} one has: 
\begin{IEEEeqnarray}{RCL}
  \label{lyapunovDerivativeSlotine1}
  \dot{V} &=&  {-}
   \begin{pmatrix}
        \dot{e}^\top & y^\top \Lambda_2 
   \end{pmatrix} \bar{K}
   \begin{pmatrix}
        \dot{e} \\  \Lambda_2 y 
   \end{pmatrix}
   {-}s^\top F_v s  {-}  e^\top \Lambda_1  K \Lambda_1 e,  \IEEEeqnarraynumspace
\end{IEEEeqnarray}
with 
\begin{IEEEeqnarray}{RCL}
\label{Kbar}
\bar{K} :=
   \begin{pmatrix}
        K & K \\ K & K
   \end{pmatrix} 
   .
\end{IEEEeqnarray}
One easily verifies that $\bar{K}$ is positive semi-definite. As a consequence, $\dot{V} \leq 0$ and the global stability of 
$(\int_0^t{e(s) \, ds},  e, s,\tilde{\pi}) = (\beta,0_n,0_n,0_p)$ follows. By using Assumption~\ref{hp:reference} and the fact that the variables 
$\int_0^t{e(s) \, ds},  e, s,$ and $\tilde{\pi}$ are bounded, one deduces that $\ddot{V}$ is bounded, which in turn implies that $\dot{V}$
is uniformly continuous. Then, the application of Barbalat's
Lemma ensures that $\dot{V}$ tends to zero. In view of~\eqref{lyapunovDerivativeSlotine1}, this implies that the error $e(t)$ converges to zero.

\subsection{Proof of Theorem~\ref{th:ExtensionSlotine}}
\label{proofExtSlotine}

Thanks to the formulation of the control result of Lemma~\ref{th:slotine}, this proof is similar to that above.
In particular, 
\begin{IEEEeqnarray}{RCL}
  \xi_c(t) = \dot{r} - \Lambda_1 e - \Lambda_2 y, \nonumber
\end{IEEEeqnarray}
with $e$ given by~\eqref{eq:errorColloc}, $y := \int_0^t{e(s) \, ds} - \beta$, and $\beta \in \mathbb{R}^m$ a properly chosen constant. 
Now, reconsider the candidate Lyapunov function~\eqref{lyapunovSlotine}. In view of 
Properties~\ref{hp:SkwSym},~\ref{hp:linearWRTparams} and the partitioning~\eqref{partitioning}, 
the application of the controls~\eqref{tauExte}-\eqref{piHatExte} renders the time derivative of $V$ as follows:
\begin{IEEEeqnarray}{RCL}
    \label{lyapunovDerivativeExt}
    \dot{V}  =  &-&s^T 
    \begin{pmatrix} Y_{\nc}(q, \dot q, \xi, \dot \xi)\hat{\pi}  \\ K s_{\coll} \end{pmatrix}
    - s^\top F_v s  \nonumber \\ &+& 2 e^\top K \Lambda_1 \dot{e} + 2 y^\top \Lambda_1 K \Lambda_2 e.  
\end{IEEEeqnarray}
Now, in view of the Property~\ref{hp:MisDP}, note that the auxiliary control input $\dot{\xi}_n$ is locally well defined since 
each leading principal minor of the mass matrix $M(q,\pi)$ is invertible when $\hat{\pi}$ lies in a neighborhood of $\pi$. As a consequence,
the choice of the auxiliary control input $\dot{\xi}_n$ in~\eqref{xiHatExte} implies that
  \begin{IEEEeqnarray}{RCL}
    \label{propertyXiDn}
       \widehat{M}_{\nc} \dot{\xi}_{\nc} + Y_{\nc} \left(q,\qD,\xi, \colvec {0_k}{\dot{\xi}_{\coll}} \right)\hat{\pi} = Y_{\nc}(q, \dot q, \xi, \dot \xi)\hat{\pi} =  
          K_{\nc} s_{\nc}.     \nonumber
 \end{IEEEeqnarray}
 In view of~\eqref{partitioning} and of the above equation, the expression of~$\dot{V}$ in~\eqref{lyapunovDerivativeExt} becomes
 \begin{IEEEeqnarray}{RCL}
    \label{lyapunovDerivativeExt1}
    \dot{V}  &{=}& {-}s_{\nc}^T K_{\nc}s_{\nc} {-} s_{\coll} K s_{\coll} 
    {-} s^\top F_v s  {+} 2 e^\top K \Lambda_1 \dot{e} {+} 2 y^\top \Lambda_1 K \Lambda_2 e. \nonumber  
\end{IEEEeqnarray}
Analogously to the proof of Lemma~\ref{th:slotine}, by substituting
\begin{IEEEeqnarray}{RCL}
  s_{\coll} = \qD_{\coll} - \xi_{\coll} = \dot{e} + \Lambda_1 e + \Lambda_2 y \nonumber
  \IEEEeqnarraynumspace
\end{IEEEeqnarray}
in the second term on the right hand side of $\dot{V}$ one obtains
\begin{IEEEeqnarray}{RCL}
  \label{lyapunovDerivativeExt2}
  \dot{V} = &-&s_{\nc}^T K_{\nc}s_{\nc} -
   \begin{pmatrix}
        \dot{e}^\top & y^\top \Lambda_2 
   \end{pmatrix} \bar{K}
   \begin{pmatrix}
        \dot{e} \\  \Lambda_2 y 
   \end{pmatrix} \nonumber \\
   &-&s^\top F_v s  -  e^\top \Lambda_1  K \Lambda_1 e \leq 0.  \IEEEeqnarraynumspace \nonumber
\end{IEEEeqnarray}
Consequently, the local stability of the equilibrium point $(\int_0^t{e(s) \, ds},  e, s,\tilde{\pi}) = (\beta,0_m,0_n,0_p)$ follows. In addition,
under the Properties~\ref{hp:MisDP}-\ref{hp:Gbounded}, Assumption~\ref{hp:reference} and the boundedness of $\dot{q}_{\nc}$, it is possible to verify that $\ddot{V}$
is bounded, which implies that $\dot{V}$ is uniformly continuous. Then, analogously to the proof of Lemma~\ref{th:slotine}, one shows that
the error $e(t)$ converges to zero.

\subsection{Proof of Lemma~\ref{lemma:desingularization}}
\label{proofDesingula}

Proof of \emph{i)}.
If $\det{\left(  \widehat{M}_{\nc} \right) } > 0$, then the matrix $\widehat{M}_{\nc}^{-1}$ exists.
By multiplying Eq~\eqref{delta} times $\hat{\pi}^\top$, one obtains:
\begin{IEEEeqnarray*}{RCL}
    \hat{\pi}^\top \delta = \sum_{i = 1}^{k}{ \hat{\pi}^\top Y_{M_{\nc}}^\top(q, e_i) \widehat{M}_{\nc}^{-1} e_i} = \sum_{i = 1}^{k}{ e_i^\top \widehat{M}^{\top}_{\nc} \widehat{M}_{\nc}^{-1} e_i} = k \text{.}
\end{IEEEeqnarray*}
Observe that $M(q, \cdot)$ is symmetric by construction. Now, since the system is underactuated, then $k \geq 1$.  Consequently, $|\delta|$ can never be zero.

Proof of \emph{ii)}.
Consider the following storage function
\begin{IEEEeqnarray*}{RCL}
    V_d := \frac{1}{2} {\det}^2(\widehat{M}_{\nc}) \text{.}
\end{IEEEeqnarray*}
It is possible to verify that the time derivative of $V_d$ is:
\begin{IEEEeqnarray*}{RCL}
    \dot{V}_d &=& {\det}^2(\widehat{M}_{\nc}) \tr\left( \widehat{M}^{-1}_{\nc} \dot{\widehat{M}}_{\nc} \right) \\
              &=& {\det}^2(\widehat{M}_{\nc}) \left[ \tr \left( \widehat{M}^{-1}_{\nc} \Upsilon \right) + \eta \delta^\top \Gamma \delta \right] \text{,}
\end{IEEEeqnarray*}
where $\eta$ as in Eq~\eqref{def:eta}, $\delta$ as in Eq~\eqref{delta} and $\Upsilon$ as defined in Eq~\eqref{def:upsilon}.
By choosing the adaptation law defined in Eq~\eqref{paramUpsExtDes}, one obtains that $\dot{V}_d \geq 0$ if $\det{ \left(\widehat{M}_{\nc} \right) } \leq \varepsilon$. As a consequence $\det(\widehat{M}_{\nc}) \geq \varepsilon$ $\forall t$ if $\det{\left(  \widehat{M}_{\nc} \right) }(0) > \varepsilon$.

\section*{ACKNOWLEDGMENT}
The authors thank Silvio Traversaro for the useful discussions during the development of this project.

\bibliographystyle{IEEEtran}
\bibliography{IEEEabrv,references}

\begin{thebibliography}{10}
\providecommand{\url}[1]{#1}
\csname url@rmstyle\endcsname
\providecommand{\newblock}{\relax}
\providecommand{\bibinfo}[2]{#2}
\providecommand\BIBentrySTDinterwordspacing{\spaceskip=0pt\relax}
\providecommand\BIBentryALTinterwordstretchfactor{4}
\providecommand\BIBentryALTinterwordspacing{\spaceskip=\fontdimen2\font plus
\BIBentryALTinterwordstretchfactor\fontdimen3\font minus
  \fontdimen4\font\relax}
\providecommand\BIBforeignlanguage[2]{{%
\expandafter\ifx\csname l@#1\endcsname\relax
\typeout{** WARNING: IEEEtran.bst: No hyphenation pattern has been}%
\typeout{** loaded for the language `#1'. Using the pattern for}%
\typeout{** the default language instead.}%
\else
\language=\csname l@#1\endcsname
\fi
#2}}

\bibitem{Olfati-Saber2000}
R.~Olfati-Saber, ``{Nonlinear Control of Underactuated Mechanical Systems with
  Application to Robotics and Aerospace Vehicles},'' Ph.D. dissertation, 2000.

\bibitem{Hera2011}
P.~L. Hera, ``{Underactuated Mechanical Systems: Contributions to trajectory
  planning, analysis, and control},'' Ph.D. dissertation, 2011.

\bibitem{Liu2013}
Y.~Liu and H.~Yu, ``{A survey of underactuated mechanical systems},''
  \emph{Control Theory \& Applications, IET}, vol.~7, no. February, pp.
  921--935, 2013.

\bibitem{Spong1998}
M.~W. Spong, ``{Underactuated Mechanical Systems},'' in \emph{Lecture notes in
  control and information sciences, Control problems in robotics and
  automation}, 1998, pp. 135--150.

\bibitem{Astrom1994}
K.~J. Astrom and B.~Wittenmark, \emph{{Adaptive Control}}, 2nd~ed.\hskip 1em
  plus 0.5em minus 0.4em\relax Boston, MA, USA: Addison-Wesley Longman
  Publishing Co., Inc., 1994.

\bibitem{Reyhanoglu1999}
M.~Reyhanoglu, ``{Dynamics and control of a class of underactuated mechanical
  systems},'' \emph{Automatic Control, \ldots}, vol.~44, no.~9, pp. 1663--1671,
  1999.

\bibitem{Brockett83}
R.~Brockett, \emph{{Asymptotic Stability And Feedback Stabilization}}, 1983.

\bibitem{Luca1996}
A.~De~Luca, R.~Mattone, and G.~Oriolo, ``{Dynamic mobility of redundant robots
  using end-effector commands},'' \emph{Robotics and Automation, \ldots}, no.
  April, pp. 1760--1767, 1996.

\bibitem{Park2009a}
M.~Park and D.~Chwa, ``{Swing-up and stabilization control of inverted-pendulum
  systems via coupled sliding-mode control method},'' \emph{Industrial
  Electronics, IEEE Transactions on}, vol.~56, no.~9, pp. 3541--3555, 2009.

\bibitem{Ghommam2010}
J.~Ghommam, F.~Mnif, and N.~Derbel, ``{Global stabilisation and tracking
  control of underactuated surface vessels},'' \emph{IET Control Theory \&
  Applications}, vol.~4, no.~1, pp. 71--88, Jan. 2010.

\bibitem{Luca2000}
A.~De~Luca and G.~Oriolo, ``{Motion Planning and Trajectory Control of an
  Underactuated Three-Link Robot via Feedback Linearization},'' \emph{Robotics
  and Automation, 2000. \ldots}, 2000.

\bibitem{Santiesteban2008}
R.~Santiesteban and T.~Floquet, ``{Second‐order sliding mode control of
  underactuated mechanical systems II: Orbital stabilization of an inverted
  pendulum with application to swing up/balancing},'' \emph{\ldots Nonlinear
  Control}, no. April 2007, pp. 544--556, 2008.

\bibitem{Spong1996}
M.~Spong, ``{Energy based control of a class of underactuated mechanical
  systems},'' \emph{1996 IFAC World Congress}, 1996.

\bibitem{GuXu}
Y.-l. Gu and Y.~Xu, ``{Under-actuated robot systems: dynamic interaction and
  adaptive control},'' \emph{Proceedings of IEEE International Conference on
  Systems, Man and Cybernetics}, vol.~1, pp. 958--963.

\bibitem{Slotine1988}
J.-J. Slotine, ``{Adaptive manipulator control: A case study},'' \emph{IEEE
  Transactions on Automatic Control}, vol.~33, no.~11, pp. 995--1003, 1988.

\bibitem{Spong1990}
M.~Spong, ``{Comments on "Adaptive Manipulator Control: A Case study"},''
  \emph{IEEE Transactions on Automatic Control}, vol.~35, pp. 761--762, 1990.

\bibitem{Siciliano2008}
B.~Siciliano and O.~Khatib, \emph{{Handbook of Robotics}}, 2008, vol.~15.

\bibitem{Khalil2004}
W.~Khalil and E.~Dombre, \emph{{Modeling, identification and control of
  robots}}.\hskip 1em plus 0.5em minus 0.4em\relax Butterworth-Heinemann, 2004.

\bibitem{Yoshida2000}
K.~Yoshida and W.~Khalil, ``{Verification of the Positive Definiteness of the
  Inertial Matrix of Manipulators Using Base Inertial Parameters},'' \emph{The
  International Journal of Robotics Research}, vol.~19, no.~5, pp. 498--510,
  May 2000.

\bibitem{Siciliano2009}
B.~Siciliano, L.~Sciavicco, L.~Villani, and G.~Oriolo, \emph{{Robotics:
  Modelling, Planning and Control}}, 2009.

\bibitem{Fumagalli2010}
M.~Fumagalli, M.~Randazzo, F.~Nori, L.~Natale, G.~Metta, and G.~Sandini,
  ``{Exploiting proximal F/T measurements for the iCub active compliance},'' in
  \emph{2010 IEEE/RSJ International Conference on Intelligent Robots and
  Systems}.\hskip 1em plus 0.5em minus 0.4em\relax Ieee, Oct. 2010, pp.
  1870--1876.

\bibitem{Fumagalli2012}
M.~Fumagalli, S.~Ivaldi, M.~Randazzo, L.~Natale, G.~Metta, G.~Sandini, and
  F.~Nori, ``{Force feedback exploiting tactile and proximal force/torque
  sensing},'' \emph{Autonomous Robots}, vol.~33, no.~4, pp. 381--398, Apr.
  2012.

\bibitem{Anderson2005a}
B.~D.~O. Anderson, ``{Failures of adaptive control theory and their
  resolution},'' \emph{Communications in Information and Systems}, vol.~5, pp.
  1--20, 2005.

\end{thebibliography}

\end{document}